\documentclass[10pt,
               showpacs, 
               amsmath,amssymb,aps,prd,
               nofootinbib,eqsecnum,amsart, superscriptaddress,
               a4paper]{revtex4}
\usepackage{epsfig}
\usepackage{graphicx,epsf}
\usepackage{color}
\usepackage{bm}
\usepackage{pgf,pgfarrows,pgfnodes,pgfautomata,pgfheaps,pgfshade}
\usepackage{amsmath,amssymb}
\def\be{\begin{equation}}
\def\ee{\end{equation}}
\def\beq{\begin{equation}}
\def\eeq{\end{equation}}

\def\bea{\begin{eqnarray}}
\def\eea{\end{eqnarray}}
\def\bi{\begin{itemize}}
\def\ei{\end{itemize}}

\def\bk{{\mathbf{k}}}
\newcommand{\fett}[1]{\boldsymbol{#1}}
\newcommand{\nab}{{\nabla}}

\definecolor{darkgreen}{rgb}{0,0.5,0}

\newcommand{\dd}{{\rm d}}
\def\H{{\cal H}}
\def\cs2{c_{\rm{s}}^2}
\newcommand\eq[1]{Eq.~(\ref{#1})}

\newcommand\A[1]{{\phi_{{#1}}}}

\def\scm{{\mathrm{scm}}}

\def\psiini{{\psi_{1{\rm ini}}}} 
\def\zetaini{\zeta_{1{\rm ini}}}
\def\etaini{\eta_{\rm ini}}
\def\etaonetai{\left(\frac{\eta}{\etaini}\right)}
\def\ncd{\newcommand}

\definecolor{contrastedorange}{RGB}{255,210,75}
\definecolor{contrastedblue}{RGB}{100,100,100}
\definecolor{contrastedgreen}{RGB}{250,0,0}
\definecolor{charcoal}{RGB}{0,0,0}
      \ncd{\rred}{\color[rgb]{0.9,0,0.2}}      
	\ncd{\ggreen}{\color[rgb]{.75,1,.3}}
	\ncd{\oorange}{\color[rgb]{1,.65,.25}}
	\ncd{\bblue}{\color[rgb]{1,1,.1}}
	\ncd{\ggrey}{\color[rgb]{0.8,0.8,0.8}}
	\ncd{\wwhite}{\color[rgb]{1,1,1}}

	\ncd{\bblack}{\color[rgb]{0,0,0}}
	\ncd{\ggreyo}{\color[rgb]{.25,.25,.25}}
	\ncd{\ggreyt}{\color[rgb]{.4,.4,.4}}
	\ncd{\ggreytt}{\color[rgb]{.45,.45,.45}}

\ncd{\CH}[1]{{\rred {\bf CH:} #1}}
\ncd{\AJC}[1]{{\oorange {\bf AJC:} #1}}

\def \beg {\begin{enumerate}}
\def \en {\end{enumerate}}

\def\M0{{\cal M}_0}

\begin{document}

\title{Second-order cosmological perturbation theory \\ and initial conditions for $N$-body simulations}

\author{Adam J.~Christopherson}
\email[]{achristopherson@ufl.edu}
\affiliation{Department of Physics, University of Florida, Gainesville, FL 32611, USA}
\author{Juan Carlos Hidalgo} \email[]{hidalgo@fis.unam.mx}
\affiliation{Instituto de Ciencias F\'{\i}sicas, Universidad Nacional Aut\'onoma de M\'exico,  Av.\ Universidad S/N, Cuernavaca, Morelos, 62251, Mexico}
\author{Cornelius Rampf}\email[]{cornelius.rampf@port.ac.uk}
\affiliation{Institute of Cosmology and Gravitation, University of Portsmouth,
Dennis Sciama Building, Burnaby Road, Portsmouth, PO1 3FX, United Kingdom}
\author{Karim~A.~Malik} \email[]{k.malik@qmul.ac.uk}
\affiliation{Astronomy Unit, School of Physics and Astronomy, Queen
  Mary University of London, Mile End Road, London, E1 4NS, United
  Kingdom}

\date{\today}

\begin{abstract}
We use gauge-invariant cosmological perturbation theory to calculate
the displacement field that sets the initial conditions for $N$-body
simulations. Using first and second-order fully relativistic
perturbation theory in the synchronous-comoving gauge, allows us to go
beyond the Newtonian predictions and to calculate relativistic
corrections to it. We use an Einstein--de Sitter model, including both
growing and decaying modes in our solutions.
The impact of our results should be assessed through the implementation of the featured displacement in cosmological $N$-body simulations.
\end{abstract}

\pacs{95.35.+d, 95.30.Sf, 98.80.-k, 98.80.Jk, 98.80.Cq \hfill } 

\maketitle

\section{Introduction}
\label{intro}

Recent years have seen tremendous progress in both observational and theoretical cosmology. The acquisition of new data allows us to test our theories and constrain parameters, enabling us to formulate the standard cosmological model. Observations of the `late' universe include maps of the large-scale structure, which have been pioneered by the Sloan Digital Sky Survey (SDSS) \cite{Eisenstein:2011sa, Dawson:2012va}. Current and future experiments including the Dark Energy Survey \cite{Abbott:2005bi}, the Large Synoptic Survey Telescope ({\sc LSST}) \cite{Ivezic:2008fe} and the {\sc Euclid} satellite \cite{Amendola:2012ys} will take even more data to build up a three dimensional image of the large-scale structure. By comparing the observed galaxy clustering with simulations, we have fairly conclusive evidence that, at the time of formation, the Universe was dominated by matter and, in turn, that the matter content was dominated by cold dark matter (CDM). 

At early times the Universe is close to being homogeneous, and the inhomogeneities that lead to the large scale structure are small in amplitude. Perturbative techniques are therefore extremely powerful tools to model the early seeds of structure formation, such as those imprinted in the cosmic microwave background. To do this, we consider the inhomogeneities as small fluctuations about a homogeneous expanding background. 

However, gravity is a nonlinear interaction, and in order to fully understand the process of structure formation, one must consider the (nonlinear) multistream regime.  To tackle this, one has to resort in general to Vlasov--Poisson solvers. Commonly used solvers are discretised $N$-body simulations which aim to approximate the phase-space dynamics to high accuracy.
The simulations themselves rely on Newtonian physics. Since the incoming data from surveys is of such high quality, we are now at a stage where relativistic corrections may become important and must be quantified. In particular, the {\sc Euclid} collaboration \cite{Amendola:2012ys} requires its $N$-body simulations to have percent-level accuracy. Relativistic corrections can be important in simulations even using Newtonian physics and cold dark matter particles, when the size of the simulated box is the same as the horizon size at the time that the initial conditions are set. At that time, effects of general relativity on such large scales could likely become important. 

Some recent work has focused on the effects of incorporating relativistic corrections into $N$-body simulations \cite{Rampf:2012pu, Adamek:2014xba, Rampf:2014mga, Valkenburg:2015dsa}, or interpreting Newtonian simulations in terms of General Relativity \cite{Rigopoulos:2014rqa,Fidler:2015npa}. Since gravity is nonlinear, the simulations are extremely sensitive to the initial conditions used. So far, the initial conditions have been specified using only linear cosmological perturbation theory \cite{mfb,ks, Bardeen:1980kt}. In this paper, we compute the second-order corrections to the initial conditions using cosmological perturbation theory and keeping the decaying mode at all orders.

The paper is structured as follows: in the next section we present the governing equations for the system for background, linear and second-order perturbations. In Section~\ref{sec:field_equations} we solve the governing field equations in the synchronous-comoving gauge, order by order. In Section~\ref{sec:displacement} we calculate the Lagrangian displacement field, and in Section~\ref{sec:discussion} we conclude.

\section{Governing equations}
\label{sect2}

In this section we present the governing equations needed to derive
the relation between the input power spectra and the displacement
fields later on. We use gauge-invariant cosmological perturbation theory,
following largely the notation of Ref.~\cite{MW2008}.

Tensorial quantities are split, here using the energy
density as an example, as
\be
\rho=\rho_0+\delta\rho_1+\frac{1}{2}\delta\rho_2\,,
\ee
up to second order. The background variables depend only on time,
whereas the perturbations depend on all four coordinates $x^\mu$.
Derivatives with respect to conformal time are denoted by a prime. 
Greek indices, $\mu,\nu,\lambda$, run from $0,\ldots 3$, while
lower case Latin indices, $i,j,k$, run from $1,\ldots3$. Einstein summation
over repeated indices is implied.

\subsection{Metric tensor}

We assume a spatially flat Friedmann--Lema\^itre--Robertson--Walker (FLRW) background, with  metric
\be
\label{def_bg_metric}
ds^2 = a^2 \left( -\dd\eta^2 + \delta_{ij} \dd x^i \dd x^j \right) \,,
\ee
where $\eta$ is conformal time, $a=a(\eta)$ the scale factor, and
$\delta_{ij}$ is the background 3-metric.
Cosmic time, measured by observers at fixed comoving spatial
coordinates, $x^i$, is given by $t=\int a(\eta)\, \dd\eta$.

The perturbed part of the metric tensor can be written as
\begin{eqnarray}
\label{defA}
\delta g_{00} &=& - 2a^2 \A{} \, , \qquad
\delta g_{0i} =   a^2 {B}_{i} \, , \qquad
\delta g_{ij} =  2a^2  C_{ij}  \, .
\end{eqnarray}
The 0-i and the i-j components of the metric tensor can be
further decomposed into scalar, vector and tensor parts
\bea
\label{decompBi}
{B}_i  &=& {B}_{,i} -S_i\, ,\\
\label{decompCij} C_{ij} &=& -\psi\;\delta_{ij} + E_{,ij}+ F_{(i,j)}
+ \frac12 h_{ij}\,.
 \eea
where $\A{}$, $B$, $\psi$ and $E$ are scalar, $S_i$ and $F_i$ are
vector, and $h_{ij}$ is a tensor metric perturbation. Note, $\A{}$ is
usually referred to as the lapse function, and $\psi$ as the
curvature perturbation.

For the results of this paper we will consider only scalar perturbations.
We will solely work in the synchronous-comoving gauge, which is
specified by the gauge conditions
\begin{align}
 \begin{split}
\phi_{1\scm}&=0\,, \qquad\, {B}_{1\scm}=0\,, \qquad\, v_{1\scm}=0\,,\\ 
\phi_{2\scm}&=0\,, \qquad\, {B}_{2\scm}=0\,, \qquad\, v_{2\scm}=0\,,
 \end{split}
\end{align}
at first and second order, where $v_{1\scm}$ and $v_{2\scm}$ are the
velocity potentials at first and second order, respectively. 
This renders the remaining first and second-order perturbations
gauge-invariant \cite{Malik:2003mv,MW2008}.

We work in an Einstein--de Sitter universe, a reasonable restriction
considering that we are interested in setting up initial conditions
for structure formation at very early times where the impact of the
cosmological constant is negligible. We however include also decaying
modes, which could have significant impact on the early gravitational
dynamics. To avoid cluttering the equations, and since there is no confusion possible, we will drop the subscript ``${\scm}$'' from now on.

\subsection{The Einstein and energy-momentum tensors}

The Einstein tensor, $G^\mu{}_{\nu}$, in the background is given by
\be
G^0_{~0}=-\frac{3}{a^2}\H^2\,, \qquad
G^0_{~i}=0\,,\qquad
G^i_{~j}=\frac{1}{a^2} \left( \H^2-2\frac{a''}{a} \right) \delta^i_{~j} \,,
\ee
where $\H\equiv a'/a$, and a prime denotes a partial time derivative with respect to conformal time.
For later convenience we also give the trace of the spatial part of the
Einstein tensor, which in the background is
\be
G^k_{~k}=\frac{3}{a^2}\left( \H^2-2\frac{a''}{a} \right)\,.
\ee
%

\subsubsection{First order Einstein tensor}

We now give the Einstein tensor in the synchronous-comoving gauge, for scalar perturbations only. 
At first order the 0-0 component of the perturbed Einstein tensor is given  by
\be
G^{0(1)}_{~0}=\frac{1}{a^2}\Bigg[{-2\nab^2\psi_1}
+2\H\left(3\psi_1'-\nab^2E_1'\right)\Bigg]\,,
\ee
the 0-i component is
\be
G^{0(1)}_{~i}=-\frac{2}{a^2}\psi_{1,i}'\,,
\ee
and the i-j component is
\be
\label{1Gij}
G^{i(1)}_{~j}=\frac{1}{a^2}\Big(
4\H\psi_1'-2\H\nab^2E_1'-\nab^2\psi_1+2\psi_1''-\nab^2E_1''
\Big)\delta^i_{~j} 
+\frac{1}{a^2}\Big( \psi_1+2\H E_1'+E_1''\Big)^{~i}_{,~j}
\,.
\ee
By contracting \eq{1Gij} with $\delta^j{}_i$ we obtain the first-order spatial trace
\be
G^{k(1)}_{~k}=\frac{1}{a^2}\Big(
12\H\psi_1'-4\H\nab^2E_1'-2\nab^2\psi_1+6\psi_1''-2\nab^2E_1''
\Big)
\,.
\ee
%

\subsubsection{Second order Einstein tensor}

At second order in the perturbations we get for the 0-0 component of
the Einstein tensor (e.g. Ref.~\cite{Christopherson:2011ra})
in the synchronous-comoving gauge
\begin{align}
\frac{1}{2}G^{0(2)}_{~0} =\frac{1}{a^2}\Bigg[
&-\nab^2\psi_2
+\H\left(3\psi_2'-\nab^2E_2'\right)
+12\H\psi_1\psi_1'-4\H\psi_1\nab^2E_1' \nonumber\\
&-4\H\psi_1'\nab^2E_1 +4\H E_{1,ij}'E_{1,}^{~ij}-3(\psi_1')^2+2\psi_1'\nab^2E_1'\nonumber \\
&+\frac{1}{2}\left(E_{1,ij}'{E_{1,}^{~ij}}'-\nab^2E_1'\nab^2E_1'\right)
{-8\psi_1\nab^2\psi_1} - 3\psi_{1,k}\psi_1^{~k}
\nonumber\\
&+2\nab^2\psi_1\nab^2E_1+2\psi_{1,k}\nab^2E_{1,}^{~k}
{+ 2E_{1,ij}\psi_{1,}^{~ij}}
\nonumber \\
&+\frac{1}{2}\left(\nab^2E_{1,k}\nab^2E_1^{~k}
-E_{1,ijk}E_{1,}^{ijk}\right)
\Bigg]\,,
\end{align}
for the 0-i component 
\begin{align}
\frac{1}{2}G^{0(2)}_{~i}=\frac{1}{a^2}\Bigg(&-\psi_{2,i}'
-4\psi_1\psi_{1,i}'-4\psi_1'\psi_{1,i}
+\nab^2 E_1' \psi_{1,i}+2\nab^2 E_1 \psi_{1,i}'
-{E_{1,}^{jk}}'E_{1,jki} \nonumber \\
&+\nab^2E_{1,}^{k}E_{1,ki}' +\psi_{1,}^{~k}E_{1,ki}'-2\psi_{1,k}'E_{1,i}^{~k}
\Bigg)\,,
\label{G0i:second}
\end{align}
and for the i-j component
\begin{align}
\frac{1}{2}&G^{i(2)}_{~j}=
\frac{1}{2a^2}\Big(
4\H\psi_2'-2\H\nab^2E_2'-\nab^2\psi_2+2\psi_2''-\nab^2E_2''
\Big)\delta^i_{~j} 
+\frac{1}{2a^2}\Big(\psi_2+2\H E_2'+E_2''\Big)^{~i}_{,~j}
\nonumber \\
&
+\frac{1}{a^2}\Big[
8\H\psi_1'\psi_1-4\H\left(\psi_1\nab^2E_1'+\psi_1'\nab^2E_1\right)
+\left(\psi_1'\right)^2+4\psi_1\psi_1''-4\psi_1\nab^2\psi_1-\psi_1'\nab^2E_1'
\nonumber \\
&
+4\H E_{1,kl}'E_{1,}^{~kl}
+\frac{3}{2} E_{1,kl}'E_{1,}^{\prime~kl}-\frac{1}{2} E_{1,klm}E_{1,}^{~klm}
+\psi_{1,k}\nab^2E_{1,}^{~k}-2\psi_1\nab^2E_1''-2\psi_1''\nab^2E_1\nonumber \\
&
+2\nab^2\psi_1\nab^2E_1+2E_{1,kl}''E_{1,}^{~kl}-2\psi_{1,k}\psi_{1,}^{~k}
+\frac{1}{2}\nab^2E_{1,k}\nab^2E_{1,}^{~k}-\frac{1}{2}\nab^2E_{1}'\nab^2E_{1}'
\Big]\delta^i_{~j} 
\nonumber \\
&
+{\frac{1}{a^2}}\Big[4\psi_1\psi_{1,~j}^{~i}+E_{1,}^{~ikl}E_{1,jkl}-2\psi_{1,~j}^{~i}\nab^2E_{1}
-\left(\psi_{1,k}+\nab^2E_{1,k}\right) E_{1,~~j}^{~ki}+2\psi_{1,}^{~ik}E_{1,kj}\nonumber \\
&
+2\psi_1 E_{1,~~j}^{\prime\prime~i}+2\left(\psi_1''+2\H\psi_1'-\nab^2\psi_1\right) E_{1,~j}^{~~i}
-2E_{1,}^{~ik}E_{1,kj}''+\left(\psi_1'+4\H\psi_1\right) E_{1,~~j}^{\prime~i}\nonumber \\
&
-2E_{1,}^{\prime~ik}E_{1,kj}' -4\H E_{1,}^{~ik}E_{1,kj}'+3\psi_{1,}^{~i}\psi_{1,j}+\nab^2E_{1}'E_{1,~j}^{\prime~i}\Big]\,.
\label{Gij:second}
\end{align}
The trace of the last expression is
\begin{align}
\label{Gkk:second}
\frac{1}{2}G^{k(2)}_{~k}
&=\frac{1}{2a^2}\Big(
12\H\psi_2'-4\H\nab^2E_2'-2\nab^2\psi_2+6\psi_2''-2\nab^2E_2''
\Big)  \nonumber \\
&
+\frac{1}{a^2}\Big[
24\H\psi_1'\psi_1-8\H\left(\psi_1\nab^2E_1'+\psi_1'\nab^2E_1\right)
+3\left(\psi_1'\right)^2+12\psi_1\psi_1''-8\psi_1\nab^2\psi_1
\nonumber \\
&\qquad\quad-2\psi_1'\nab^2E_1' +8\H E_{1,kl}'E_{1,}^{~kl}
+\frac{5}{2} E_{1,kl}'E_{1,}^{\prime~kl}-\frac{1}{2} E_{1,klm}E_{1,}^{~klm}
+2\psi_{1,k}\nab^2E_{1,}^{~k} \nonumber \\
&\qquad\quad -4\psi_1\nab^2E_1''-4\psi_1''\nab^2E_1
+2\nab^2\psi_1\nab^2E_1+4E_{1,kl}''E_{1,}^{~kl}-3\psi_{1,k}\psi_{1,}^{~k} \nonumber \\
&\qquad\quad+\frac{1}{2}\nab^2E_{1,k}\nab^2E_{1,}^{~k}-\frac{1}{2}\nab^2E_{1}'\nab^2E_{1}'
+2\psi_{1,jk}E_{1,}^{~jk}
\Big]  \,.
\end{align}
%

\subsubsection{The energy-momentum tensor}

As mentioned above, for simplicity we assume a pressureless perfect fluid with vanishing anisotropic stress and work in the synchronous-comoving gauge.
For this setup, the time-time component of the energy-momentum tensor, $T^\mu{}_{\nu}$, is, up to second order,
\begin{align}
{}^{(0)}T^0{}_{0} &= -\rho_0 \,, \qquad \quad
{}^{(1)}T^0{}_{0} = - \delta \rho_1\,,  \qquad \,\,
{}^{(2)}T^i{}_{j} = \delta \rho_2 \,.
\end{align}
The rest of the components are zero up to second order in perturbations. Note the significant simplifications in the expression for the perturbed energy-momentum tensor compared to the general form (see
e.g.~Ref.~\cite{MW2008}) that stem from the matter content and the
choice of gauge.

\subsection{Energy and momentum conservation}

The energy-momentum conservation equations are given by
\be
\nabla_\mu T^\mu{}_{\nu}=0\,,
\ee
where $\nabla_\mu$ is the covariant derivative.  In the background,
for the considered fluid, the $\nu=0$ component from the above gives
\begin{align}
\label{continuity:back}
\rho_0'+3\H\rho_0&=0\,,
\end{align}
at first order,
\begin{align}
\label{continuity:first}
\delta\rho_{1}'+3\H\delta\rho_1
-3\rho_0\psi_1'
+ \rho_0\nab^2 E_1'  &=0\,,
\intertext{and at second order} 
\delta\rho_{2}'+3\H\delta\rho_2
-3\rho_0\psi_2'+ \rho_0\nab^2 E_2'
-6\psi_1' \delta\rho_1
+2\delta\rho_1\nab^2 E_1' &
\nonumber\\
=2\rho_0\Big(
6\psi_1\psi_1'-2\psi_1\nab^2 E_1'-2\psi_1'\nab^2 E_1+&2E_{1,ij}'E_{1,}^{~ij}
\Big) \,.
\label{continuity:second}
\end{align}
It is also often convenient to work in terms of the density contrast,
$\delta$, which is defined as
\be
\label{def_dens_cont}
\delta_1=\frac{\delta\rho_{1}}{\rho_0}\,, \qquad 
\delta_2=\frac{\delta\rho_{2}}{\rho_0}\,,
\ee
at first and second order, respectively.

\subsection{Field equations}

The Einstein field equations, governing the dynamics of spacetime, are 
\be
\label{fieldequations}
G^\mu{}_\nu=8\pi G\, T^\mu{}_\nu\,,
\ee
where $G$ is Newton's gravitational constant, and we have set the speed of light to unity.
To reiterate, we consider an Einstein--de Sitter universe, i.e., one filled by pressureless dust and vanishing cosmological constant. The dust is modelled as a single and perfect fluid, which amounts to a continuum description for collisionless dark matter, valid before the first instance of shell-crossing.

\subsubsection{Background}

The 0-0 component of \eq{fieldequations} in the background yields the Friedmann equation
\be \label{Friedmann}
\H^2=\frac{8 \pi G}{3}\rho_0 a^2\,.
\ee
At zeroth order in perturbation theory, this is the only equation we require in addition to the continuity equation~(\ref{continuity:back}).

\subsubsection{First order}

With the assumptions on the matter content given above, namely no
pressure and no anisotropic stress, we have the following governing
equations in the synchronous-comoving gauge at first order,
\begin{align}
\label{curvature1}
\psi_{1}'&=0\,,\\
\label{shear1}
\sigma_{1}'+2\H\sigma_{1}+\psi_{1} &=0\,.
\end{align}
In an arbitrary gauge, the scalar shear is defined as $\sigma_{1}\equiv-B_1+E_1'$.  This allows us to write the evolution equation for $\delta\rho_1$ as
\be
\label{densitycont1}
\delta\rho_{1}'+3\H\delta\rho_{1}+\rho_0\nab^2\sigma_{1}=0\,.
\ee
We find, however, more useful to rewrite \eq{shear1} in terms of the metric function $E_{1}$. Using the definition of the shear scalar in the synchronous-comoving gauge, 
\be
\label{shear1E1}
\sigma_{1}\equiv E_{1}'\,,
\ee
we obtain, 
\be 
\label{eq:E1}
E_{1}'' + 2 {\cal H} E_{1}' = - \psi_{1} \,.
\ee
%

\subsection{Second order}

We now write the governing equations at second order.
From integrating in time the divergence of \eq{G0i:second},
we get 
\be
\label{laplacian:psi2}
\nab^2 \psi_2 = \nab^2 \psi_1 \nab^2 E_1 + 2\psi_{1,i}\nab^2 E_{1,}^{~i} 
+ \psi_{1,ki} E_{1,}^{~ki} 
+ \frac{1}{2} \nab^2 E_{1,k}\nab^2 E_{1,}^{~k} 
- \frac{1}{2} E_{1,ijk}E_{1,}^{~ijk} \,.
\ee 
{}From the trace of the i-j component of the field equations, \eq{Gkk:second} above, we
find an evolution equation for $\nab^2E_2'$, that is
\begin{align}
\label{trace_2nd}
\nab^2E_2''+2\H \nab^2E_2' &= 3\nab^{-2}\Bigg[
 -\nab^2\psi_1 \nab^2 \psi_1 
- \nab^2 \psi_{1,k} \left(2\psi_{1,}^{~k}+\nab^2E_{1,}^{~k}\right) \nonumber\\
&\qquad\quad
- \psi_{1,kl}\psi_{1,}^{~kl}+ \psi_{1,ijk} E_{1,}^{~ijk}
+\nab^2E_{1,k}'\nab^2E_{1,}^{\prime~k}-E_{1,ijk}'E_{1,}^{\prime~ijk}\Bigg] \nonumber\\
-4\psi_1 \nab^2 \psi_1 - &3\psi_{1,k}\psi_{1,}^{~k}
-3E_{1,}^{~kl} \psi_{1,kl} 
+\frac{5}{2}E_{1,kl}'E_{1,}^{\prime~kl} - \frac{1}{2}\nab^2E_{1}'\nab^2E_{1}'
+\nab^2E_{1}\nab^2\psi_{1}\,,
\end{align}
where we have used Eq.\,(\ref{eq:E1}), and \eq{laplacian:psi2} and its time derivative. Note that
\eq{trace_2nd} is only assuming dust and neither assuming growing mode only
nor Einstein--de Sitter.

\section{Solutions to the field equations}
\label{sec:field_equations}

In this section we solve the governing evolution equations as given in the above section, order by order.
Here we shall report only the solutions in real space, for results in Fourier space see Appendix \ref{app:Fourier}.

\subsection{Background}

Equations~(\ref{continuity:back}) and~(\ref{Friedmann}) yield the
well known dust solutions
\be
 \rho_0\propto a^{-3}\,,\qquad   a\propto \eta^2\,, \qquad \H=\frac{2}{\eta} \,.
\ee

\subsection{First order}

At linear order we find that the
solutions to the system of governing equations, Eqs.\,(\ref{curvature1})--(\ref{densitycont1}), are given by
\bea
\psi_{1}&=&\psiini(x^i)\,,\\
\sigma_{1}(\eta,x^i)
&=&\sigma_{1{\rm ini}}\left(\frac{\eta_{\rm ini}}{\eta}\right)^4
+\frac{1}{5}\psiini(x^i)\eta_{\rm ini}\Big[\left(\frac{\eta_{\rm ini}}{\eta}\right)^4
-\frac{\eta}{\eta_{\rm ini}} \Big]\,.\\
\delta\rho_{1}(\eta,x^i)&=&\frac{\nab^2}{4 \pi G a^2} 
\Big[\psiini(x^i)+\frac{2}{\eta}\sigma_{1}(\eta,x^i)\Big]\,,
\eea
where the spectrum of the curvature perturbation initially is
$\psiini(x^i)$ and we set $\sigma_{1{\rm ini}}\equiv \sigma_{1}(\eta_{\rm
  ini},x^i)$.
Hence the dominant contribution to the shear at late times is
\be
\sigma_{1}(\eta,x^i)\simeq -\frac{\eta}{5}\psiini(x^i)\,.
\ee

We also need the metric function $E_1$, which is simply the time
integrated shear, as defined in \eq{shear1E1}. The curvature
perturbation on uniform density hypersurfaces is defined, see
e.g.~Ref.~\cite{MW2008, Wands:2000dp}, as
\be
\label{defzeta1}
\zeta_1=-\psi-\frac{\H}{\rho_0'}\delta\rho_1\,,
\ee
where the RHS of \eq{defzeta1} is gauge invariant by construction and
can therefore also be evaluated in synchronous-comoving gauge. For
dust $\zeta_1$ simplifies to $\zeta_1 = - \psi_1 +  \delta_1/3$, a constant in time. We will use this variable in the following as a convenient quantity to specify the second initial condition in the solution for $E_1$, besides $\psiini(x^i)$.
We then get the solution to \eq{eq:E1} as
\beq
\label{E1solutionfull}
E_1(\eta,x^i) = - \frac{1}{10} \left(\eta^2 - \etaini^2\right) \psiini(x^i) 
+ 3\left(\frac{\eta}{\etaini}\right)^{-3}\nabla^{-2} \left[\zetaini(x^i) 
+\psiini(x^i)\right]\,,
\eeq
where the initial condition for $E_1$, at the initial time $\etaini$, is simply
the pure decaying mode,
\beq
E_1(\etaini,x^i) = 3 \nab^{-2} \left[\zetaini(x^i) +\psiini(x^i)\right]\,.
\eeq

\noindent as is also the initial condition for the shear:
\beq
\sigma_{1 {\rm ini}} = - \frac{\etaini}{5} \left[\psiini(x^i) + 45 \etaini^{-2}\,\nab^{-2}\left(\psiini(x^i) + \zetaini(x^i)\right)\right]\,.
\eeq

With the above results we compute the full first order matter density contrast. We integrate Eq.~\eqref{densitycont1} to find
\be
\delta_1 = -\nab^2E_1 =  \frac{1}{10} \eta^2  \nab^2 \psiini(x^i) 
- 3\left(\frac{\eta}{\etaini}\right)^{-3} \left[\zetaini(x^i) 
+\psiini(x^i)\right]\,, 
\ee

Note that the dominant contributions at late times to $E_1$ and $\delta_1$, or the growing modes, are thus
\bea
E_1(\eta,x^i)&=& -\frac{\eta^2}{10}\psiini(x^i)\,,\qquad
\delta_1 = \frac{\eta^2}{10}\nab^2 \psiini(x^i)\,.
\label{growing:mode}
\eea
%

\subsection{Second order}

\subsubsection{Growing mode only}
\label{grow_sol_sect}

We now give the solutions for the second-order metric functions $\psi_2$, $E_2$, 
and for the density contrast $\delta_2$, taking only the growing mode
solutions above into account. Here we present them in real space, for the solutions in  Fourier space, see Appendix~\ref{app:Fourier}. The solutions including the decaying mode are given in the following subsection.

From Eq.~(\ref{laplacian:psi2}) and using the first-order results for
$\psi_1$ and $E_1$, we obtain after some simple manipulations the fastest-growing mode solution
\be
\label{psi2:CR}
\psi_2  = - \frac{\eta^2}{10} \left( \psi_{1,m} \psi_1^{,m} 
+  \nab^{-2} \left[ (\nab^2\psi_1)^2- \psi_{1,lm}\psi_1^{,lm} \right]   \right)  
+ \frac{\eta^4}{200} \nab^{-2} \left( \psi_{1,lm}^{,l} \psi_{1,k}^{,km}
      - \psi_{1,klm} \psi_1^{,klm} \right) \,.
\ee

\noindent Hereafter we omit the subscript $\mathrm{ini}$ in $\psi_1$ and $\zeta_1$ for simplicity, but keeping in mind that both perturbations are time-independent. To obtain the fastest-growing mode solution for $E_2$ we first plug in the first-order solutions in the differential equation~(\ref{trace_2nd}) for $E_2$. Then, multiplying this differential equation by $\nab^{-2}$ we obtain
\begin{align} \label{simpleODE_E2}
 E_2''+2\H E_2' = 6 \Theta_0 - 2 \psi_1^2 +  \frac{\eta^2}{25} \nab^{-2} \Bigg[ &
    \frac{21}{2} \nab^{-2} \left( \nab^2 \psi_{1,k} \nab^2 \psi_{1}^{~,k}   -  \psi_{1,klm} \psi_{1,}^{~klm} \right) \nonumber \\
 & + 10 \psi_{1,kl} \psi_1^{,kl} - 3 (\nab^2 \psi_1)^2 
\Bigg] \,,
\end{align}
where we have defined the kernel
\begin{align}
\nab^2 \Theta_0 &= - \nab^{-2} \mu_2 - \frac 1 3 \psi_{1,k} \psi_1^{,k} \,, \label{Theta0} \\
\intertext{and}
 \mu_2 &= \frac 1 2 \left[ (\nab^2\psi_1)^2- \psi_{1,lm}\psi_1^{,lm} \right] \,. \label{mu2}
\end{align}
The fastest growing solution of~(\ref{simpleODE_E2}) is then easily obtained, it is
\begin{align} 
\label{solE2}
  E_2 = \frac{3\eta^2}{5} \left( \Theta_0 - \frac 1 3 \psi_1^2 \right)
    +  \frac{\eta^4}{700} \nab^{-2} \Bigg[ &
    \frac{21}{2} \nab^{-2} \left( \nab^2 \psi_{1,k} \nab^2 \psi_{1}^{~,k}   -  \psi_{1,klm} \psi_{1,}^{~klm} \right) \nonumber \\
 &+ 10 \psi_{1,kl} \psi_1^{,kl} - 3 (\nab^2 \psi_1)^2 
\Bigg] \,.
\end{align}

The evolution equation for the second-order density contrast,
$\delta_2$, follows from the continuity equation for $\delta \rho_2$,
\eqref{continuity:second}, and using the definition
(\ref{def_dens_cont}). We get
\be
\label{delta2'}
\delta_2' =  3 \psi_2' - \nab^2E_2' -2 \delta_1 \nab^2 E_1' 
- 4\left[\psi_1 \nab^2 E_1' - E_{1,ij}' E_{1,}^{~ij}\right]\,.
\ee
Using the above results, we then obtain for the fastest-growing mode  of the second-order density contrast
\be \label{delta2+}
\delta_2 
= \frac{\eta^2}{5} \left( \frac{3}{2} \psi_{1,m} \psi_1^{,m}  + 4  \psi_1 \nab^2 \psi_1  
\right)
+ \frac{\eta^4}{50} \left( \frac{5}{7} (\nab^2\psi_1)^2 
+ \frac 2 7 \psi_{1,kl} \psi_1^{,kl} \right)  \,.
\ee
This result agrees with Ref.~\cite{Rampf:2014mga} for their $f_{\rm nl}=-5/3$,
 and agrees with Ref.~\cite{Villa:2015ppa} for their $a_{\rm nl}=0$.

\subsubsection{Growing and decaying mode}
\label{grow_and_decay_sol_sect}

We now give the solutions for the second-order metric functions $\psi_2$, $E_2$, 
including the growing and decaying modes, that is we use \eq{E1solutionfull}. 
We follow the same steps to calculate the solutions as described in
the previous section. To arrive at the expressions below, we have used the identity 
\be
\label{identity}
\left( \nab^{-2} \partial^k \partial^l - \delta^{kl} \right) \psi_{1,km} \psi_{1,l}^{,m}
     = \nab^{-2} \left( \nab^2\psi_{1,m} \nab^2 \psi_{1}^{,m}
      - \psi_{1,klm} \psi_1^{,klm} \right) \,,
\ee

\noindent and its equivalent for $(\psi_1 + \zeta_1)$. We find for $\psi_2$,
\begin{align}
\notag
 &\psi_2 = - \frac{(\eta^2- \etaini^2)}{10} \left( \psi_{1,m} \psi_1^{,m} +  2 \nab^{-2} \mu_2   \right)  + \frac{(\eta^2 - \etaini^2)^2}{200} \nab^{-2} \left( \psi_{1,lm}^{,l} \psi_{1,k}^{,km}
      - \psi_{1,klm} \psi_1^{,klm} \right) \nonumber \\ 
&+3 \left(\frac{\eta}{\etaini}\right)^{-3} \left[\nab^{-2} (\psi_{1,j} \psi_{1,}^{~j}) + 
\frac12 \left(\psi_1^2 + \nab^{-2} \psi_{1,jk} \nab^{-2}\psi_{1,}^{~jk}\right)\right] \nonumber \\
&-\frac{3}{10}(\eta^2- \etaini^2) \etaonetai^{-3} \left[\nab^{-2} \left(\psi_{1,j}\nab^2\psi_{1,}^{~j}\right) - \frac12 \nab^{-2}\psi_{1,ijk}\nab^{-2}\psi_{1,}^{~ijk} \right] \nonumber \\
&+\frac92 \left(\frac{\eta}{\etaini}\right)^{-6} \nab^{-2}\left[\psi_{1,j} \psi_{1,}^{~j} - 
 \nab^{-2} \psi_{1,ijk} \nab^{-2}\psi_{1,}^{~ijk}\right] \nonumber \\
&-3  \left(\frac{\eta}{\etaini}\right)^{-6}\nab^{-2}\left[2 \psi_{1,i}\zeta_{1,}^{~i} - \psi_{1,jk} \nab^{-2}\zeta_{1,}^{~jk}\right] 
+ \frac92\left(\frac{\eta}{\etaini}\right)^{-6} \nab^{-2}\left[\zeta_{1,j} \zeta_{1,}^{~j} - 
 \nab^{-2} \zeta_{1,ijk} \nab^{-2}\zeta_{1,}^{~ijk}\right]  \nonumber \\
&-\frac3{10}(\eta^2- \etaini^2) \etaonetai^{-3} \nab^{-2} \left[\nab^2\psi_{1,j}\zeta_{1,}^{~j} -  \psi_{1,ijk}\nab^{-2}\zeta_{1,}^{~ijk} \right] \nonumber \\
&+9  \left(\frac{\eta}{\etaini}\right)^{-3}\nab^{-2}\left[\psi_{1,i}\zeta_{1,}^{~i} - \nab^{-2}\psi_{1,ijk}\nab^{-2} \zeta_{1,}^{~ijk}\right] \,.
\label{psi2:decay}
\end{align}
The growing plus decaying solution for $E_2$ is derived from expanding Eq.~\eqref{trace_2nd} in terms of the pure growing mode $\psi_1$ and the decaying mode  initially given by $E_1$ in Eq.~\eqref{E1solutionfull}:
\begin{align}  
\notag
&E_2''+2\H E_2' = 6 \Theta_0 - 2 \psi_1^2  +24\left( \etaonetai^{-3} + \frac34 \etaonetai^{-8}\right) \nab^{-2}\left[\psi_{1,kl}\nab^{-2}(\zeta_1+ \psi_1)_{,}^{~kl}\right] \nonumber \\ 
 & +\nab^{-2} \Bigg\{ \frac{\eta^2}{25}
    \frac{21}{2}\left(1 +\frac17 \etaonetai^{-5}+ \frac{9}{14} \etaonetai^{-10}\right) \nab^{-2} \left[ \nab^2 \psi_{1,k} \nab^2 \psi_{1}^{~,k}   -  \psi_{1,klm} \psi_{1,}^{~klm} \right]  \nonumber \\
 &+\frac{243}{\eta^2} \etaonetai^{-6}\nabla^{-2}\left[(\zeta_1+\psi_1)_{,k}(\zeta_1+\psi_1)^{,k} - \nab^{-2}(\zeta_1+\psi_1)_{,ijk}\nab^{-2}(\zeta_1+\psi_1)^{,ijk}\right]
\nonumber \\
&+  \frac95 \left(\etaonetai^{-3}+ 4 \etaonetai^{-8}\right)\nabla^{-2}\left[(\zeta_{1} + \psi_1)_{,k} \nab^2\psi_{1,}^{~k} + \nab^{-2}(\zeta_1+\psi_1)_{,ijk}\psi_{1,}^{~ijk}\right]
 \nonumber \\
&+\frac{10}{25}\eta^2\left(1 + \frac{9}{16} \etaonetai^{-10}\right)\psi_{1,ij}\psi_{1,}^{~ij}  + \frac65 \left( \etaonetai^{-3} + \etaonetai^{-8}\right) \left(\zeta_1+\psi_1\right) \nab^2 \psi_1 \nonumber \\
&-\frac{3}{25}\eta^2\left(1 - \frac13 \etaonetai^{-5} - \frac38 \etaonetai^{-10}\right) \nab^2\psi_{1}\nab^2\psi_{1,}  \nonumber \\
&+ \frac{81}{2} \eta^2 \etaonetai^{-10}\left[ 5\nab^{-2}(\zeta_1+ \psi_1)_{,kl} \nab^{-2}(\zeta_1+ \psi_1)_,^{~kl}- (\zeta_1+ \psi_1)^2 \right]\Bigg\}
 \,.  \label{decay:E2ODE}
\end{align}
Solving the ODE for $E_2$ we obtain the following expression 
\begin{align}
&E_2 = \frac{3}{5}\eta^2 \left(\Theta_0 - \frac 1 3 \psi_1^2\right) 
- \frac35\eta^2 \left(\etaonetai^{-3} - \frac19\etaonetai^{-8}\right) \nab^{-2}\left[\left(\zeta_1+\psi_1\right) \nab^2 \psi_1  \right] 
  \nonumber \\ 
&+ \nab^{-2} \Bigg\{ \frac{\eta^4}{70} \left[ 1 + \frac{7}8 \etaonetai^{-10}\right] \psi_{1,ij}\psi_{1,}^{~ij}  
-\frac{3\eta^4}{700} \left[ 1 - \frac{14}{3} \etaonetai^{-5} - \frac7{12} \etaonetai^{-10}\right] \nab^2\psi_{1}\nab^2\psi_{1}   \nonumber \\
 &+\frac32 \frac{\eta^4}{100} \left(1 - 2 \etaonetai^{-5}+  \etaonetai^{-10}\right)  
 \nab^{-2} \left[ \nab^2 \psi_{1,k} \nab^2 \psi_{1}^{~,k}   -  \psi_{1,klm} \psi_{1,}^{~klm} \right]\nonumber\\
&- \eta^2\left(  12\etaonetai^{-3} - \etaonetai^{-8}\right)\psi_{1,kl}\nab^{-2}(\zeta_1+ \psi_1)_{,}^{~kl}  \nonumber\\
&-  \frac9{10}\eta^2 \left(\etaonetai^{-3}- \frac49\etaonetai^{-8}\right) \nabla^{-2}\left[(\zeta_{1} + \psi_1)_{,k} \nab^2\psi_{1,}^{~k} + \nab^{-2}(\zeta_1+\psi_1)_{,ijk}\psi_{1,}^{~ijk}\right] \nonumber \\
&+\frac{27}{2} \etaonetai^{-6}\nabla^{-2}\left[(\zeta_1+\psi_1)_{,k}(\zeta_1+\psi_1)^{,k} - \nab^{-2}(\zeta_1+\psi_1)_{,ijk}\nab^{-2}(\zeta_1+\psi_1)^{,ijk}\right] \nonumber \\
&+ \frac{9}4  \etaonetai^{-10}\left[ 5\nab^{-2}(\zeta_1+ \psi_1)_{,kl} \nab^{-2}(\zeta_1+ \psi_1)_,^{~kl}- (\zeta_1+ \psi_1)^2 \right] \Bigg\}
 \,.  \label{decay:E2}
\end{align}

For completeness we also give the second order density contrast
including growing and decaying modes,
\begin{align}
&\delta_2=\frac3{10} \left(\eta^2 +\etaini^2\right) \psi_{1,k}\psi_{1,}^{~k}+ \frac{4}{5} \left(\eta^2 +\etaini^2\right) \psi_1 \nab^2 \psi_1 \nonumber \\
&+ \frac{\eta^4}{100} \left[\frac{10}{7} - 2 \etaonetai^{-2} + \etaonetai^{-4} - 2 \etaonetai^{-5} - \frac14 \etaonetai^{-10}   \right] \nab^2\psi_1\nab^2 \psi_1\nonumber \\
&+ \frac{\eta^4}{100} \left[\frac{4}{7} - 4 \etaonetai^{-2} + 2\etaonetai^{-4} + 3 \etaonetai^{-5} - \frac{11}{4} \etaonetai^{-10}   \right] \psi_{1,ik}\psi_{1,}^{~ik}\nonumber \\
&+ \frac{3}{100}\etaini^4 \left[\frac{1}{2} - 2 \etaonetai^{-2} + 2 \etaonetai^{-3} + \etaonetai^{-4}  - \etaonetai^{-10}   \right] \left(\nab^{-2}(\nab^2\psi_{1,k}\nab^2 \psi_{1,}^{~k}) - \nab^{-2}(\psi_{1,ikj}\psi_{1,}^{~ijk}) \right)\nonumber \\
&- \frac35 \etaini^2 \nab^{-2}\left(\psi_{1,ik}\psi_{1,}^{~ik}\right) 
- \frac{13}{2} \etaonetai^{-3}(\psi_1 + \zeta_1)\psi_1 + 9 \left[\etaonetai^{-6}+ \frac14 \etaonetai^{-10}\right](\psi_1+\zeta_1)^2\nonumber  \\
&+ \frac35 \etaini^2 \left[\etaonetai^{-3}- \frac19 \etaonetai^{-6}\right] (\psi_1 + \zeta_1)\nab^2\psi_1 
+ 9 \etaonetai^{-3}\nab^{-2} \left(\psi_{1,k}(\psi_1 + \zeta_1)_{,}^{~k}\right) \nonumber \\
& + \etaini^2\left[\frac{123}{10} \etaonetai^{-1} - \frac{3}{10}\etaonetai^{-3} - \etaonetai^{-6}\right] \psi_{1,ik}\nab^{-2}(\psi_1 + \zeta_1)_{,}^{~ik} \nonumber \\
&+ 9 \etaonetai^{-3}\nab^{-2} \left(\psi_{1,ik}\nab^{-2}(\psi_1 + \zeta_1)_{,}^{~ik}\right) - \frac95 \etaini^2\left[\etaonetai^{-1} - \etaonetai^{-3}\right] \nab^{-2} \left(\nab^{-2}(\psi_1+ \zeta_1)_{,lm}\psi_{1,}^{~km}\right)_{,k}^{~l}\nonumber \\
& + 9 \left[2\etaonetai^{-6} - \frac54 \etaonetai^{-10} \right] \nab^{-2} (\psi_1 + \zeta_1)_{,kl}\nab^{-2}(\psi_1 + \zeta_1)_{,}^{~kl}\nonumber\\
& + \frac9{10}\etaini^{4} \left[\etaonetai^{-1} - \frac49 \etaonetai^{-6}\right] \nab^{-2}\Big((\psi_1+ \zeta_1)_{,k}\nab^2(\psi_1)_{,}^{~k} - \nab^{-2}(\psi_1+ \zeta_1)_{,ijk}(\psi_1)_{,}^{~ijk}\Big) \nonumber \\
& + \frac{27}{2} \etaonetai^{-6}\Big((\psi_1+ \zeta_1)_{,k}(\psi_1+ \zeta_1)_{,}^{~k} - \nab^{-2}(\psi_1+ \zeta_1)_{,ijk}\nab^{-2}(\psi_1+ \zeta_1)_{,}^{~ijk}\Big) 
\label{delta2:gd}
\end{align}

\section{Lagrangian displacement field} 
\label{sec:displacement}

Given the solutions for the metric potentials in the previous sections, we
can now determine the metric tensor for scalar perturbations, 
and use this to calculate the corresponding Lagrangian displacement field.

Let us briefly outline the used method to obtain the displacement
field. It requires the knowledge of the metric tensor which, after
having established the above results for the metric coefficients, can
now be easily constructed.  The displacement field can then be
extracted from the metric tensor by using a suitable decomposition,
which we also motivate briefly in the following (for further details
see Ref.~\cite{Rampf:2014mga}).

We begin by writing down the metric tensor $\gamma_{ij}$ in
synchronous-comoving gauge, defined by the line element
\be
\dd s^2 = a^2(\tau) \left[ - \dd \tau^2 + \gamma_{ij}\, \dd q^i \dd q^j \right] \,,
\ee
where the $q^i$ are Lagrangian coordinates and thus denote space-like
labels of fluid elements on a time-like hypersurface (we assume the
vanishing of the spatial vorticity). In this gauge
\be
\gamma_{ij}(\tau,q^m) = \delta_{ij}\left(1 - 2 \psi(\tau, q^m)\right) + 2 E(\tau, q^m)_{, ij} 
\ee

Let us now turn to a Lagrangian description.
Let $\fett{q} \mapsto \fett{x}(\eta,\fett{q})$ be the (spatial)
Lagrangian map from the initial position $\fett{q}$ of the particle to
the final position $\fett{x}$ at time $\eta$. The respective Jacobian
matrix element is  (we neglect vectors)
\be
   {\cal J}_{ij}(\eta,\fett{q}) \equiv \delta_{ij} + {\cal F}_{,ij}(\eta,\fett{q}) \,,
\ee
where ${\cal F}$ is the scalar part of the displacement, and ``$,i$'' denotes, as before, 
a partial derivative with respect to Lagrangian coordinates.
With these definitions we can utilize the following decomposition for the spatial metric
\be
\label{gamma:displacement}
  \gamma_{ij} = \delta_{ab} \left( 1 - 2 {\cal B}  \right) \, {\cal J}^a{}_i \, {\cal J}^b{}_j \,,
\ee
with ${\cal B}= {\cal B}_1 +{\cal B}_2/2 +\ldots$ and ${\cal F}= {\cal F}_1 + {\cal F}_2/2 + \ldots$ to be determined.
We show in the following two subsections how to derive these unknowns, order by order.

\subsection{First order}

The spatial metric for scalar perturbations is, substituting the above
solutions,
\bea 
\label{gamma1}
\gamma_{1ij}&=& \delta_{ij}\left( 1 -2\psi_1 \right)
-
2\frac{\eta_{\rm ini}}{3}\sigma_{1{\rm ini},ij}\left(\frac{\eta_{\rm ini}}{\eta}\right)^3
-\frac{\eta^2}{5}\psi_{1,ij}
\Big[1+\frac{2}{3}\left(\frac{\eta_{\rm ini}}{\eta}\right)^5\Big] \,.\\
&=& \delta_{ij} \left(1 - 2 \psi_1\right) - \frac15 \left(\eta - \etaini\right)^2 \psi_{1,ij} + 6 \etaonetai^{-3} \nab^{-2} \left(\psi_1 + \zeta_1\right)_{,ij} 
\eea
At late times, or using only the growing mode, this simplifies to
\be
\label{pert_zeld1}
\gamma_{1ij}^+\simeq \delta_{ij}\left( 1-2\psi_1 \right)
-\frac{\eta^2}{5}\psi_{1,ij} \,.
\ee
To obtain the first-order displacement field, we note that the
decomposition~(\ref{gamma:displacement}) reduces at first order to
\be
\gamma_{1ij} =  \delta_{ij} \left( 1 - 2{\cal B}_1 \right) + 2 {\cal F}_{1,ij} \,,
\ee
and thus we can easily read-off the displacement by using our result
for $\gamma_{ij}$.  For the fastest growing mode, i.e.,
using~(\ref{pert_zeld1}), we obtain at first order
\begin{align}
  {\cal B}_1 &= \psi_1 \,, \\
  {\cal F}_1^+ &=  - \frac{\eta^2}{10} \psi_1 \,,  \label{F1+}
\end{align}
whereas, including the decaying modes (using Eq.\,(\ref{gamma1})) we
find the same result for ${\cal B}_1$ but for the (scalar component of the)
first-order displacement we have now
\begin{align} \label{FOdisp}
 {\cal F}_1 &=  E_1 = -\frac{\eta_{\rm ini}}{3}\sigma_{1{\rm ini}}\left(\frac{\eta_{\rm ini}}{\eta}\right)^3 -\frac{\eta^2}{10}\psi_1 \Big[1+\frac{2}{3}\left(\frac{\eta_{\rm ini}}{\eta}\right)^5\Big]\,,\nonumber \\
& = -\frac1{10} \left( \eta^2 - \etaini^2\right) \psi_1 + 3 \etaonetai^{-3}\nab^{-2}\left(\psi_1 + \zeta_1 \right)\,.
\end{align}
Equation~(\ref{F1+}) is nothing but the Newtonian displacement in the Zel'dovich approximation, whereas Eq.\,(\ref{FOdisp}) is its general relativistic extension including decaying modes. To our knowledge, the laster is a new result. Clearly, at late times the relativistic displacement~(\ref{FOdisp}) coincides with the Newtonian fastest growing mode as it should.

\subsection{Second order}

\subsubsection{Growing mode only}

We can now give the metric up to second order.
We begin with the metric including only the growing
mode solutions for the metric potentials given in Section~\ref{grow_sol_sect}.
Putting everything together the metric 
up to second order is
\begin{align}
\gamma_{ij}  &= \delta_{ij} \left[  1 - 2 \psi_1 - \psi_2 \right] + 2 E_{1,ij} + E_{2,ij} \\ 
       &= \delta_{ij} \left[ 1- 2 \psi_1 + \frac{\eta^2}{10} \left( \psi_{1,m} \psi_1^{,m} +  2 \nab^{-2} \mu_2   \right)  - \frac{\eta^4}{200} \nab^{-2} \left( \nab^2\psi_{1,m} \nab^2 \psi_{1}^{,m}
      - \psi_{1,klm} \psi_1^{,klm} \right)
 \right] \nonumber \\
  & - \frac{\eta^2}{5} \psi_{1,ij} 
   + \frac{3\eta^2}{5} \left( \Theta_0 - \frac 1 3 \psi_1^2 \right)_{,ij} \nonumber \\
   &+  \frac{\eta^4}{700} \nab^{-2} \left[ 
    \frac{21}{2} \nab^{-2} \left( \nab^2 \psi_{1,k} \nab^2 \psi_{1}^{~,k}   -  \psi_{1,klm} \psi_{1,}^{~klm} \right)
 + 10 \psi_{1,kl} \psi_1^{,kl} - 3 (\nab^2 \psi_1)^2 
\right]_{,ij}  \nonumber  \\ 
       &= \delta_{ij} \left[ 1- 2 \psi_1 + \frac{\eta^2}{10} \left( \psi_{1,m} \psi_1^{,m} +  2 \nab^{-2} \mu_2   \right) 
 \right] \nonumber \\
  & - \frac{\eta^2}{5} \psi_{1,ij} 
   + \frac{3\eta^2}{5} \left( \Theta_0 - \frac 1 3 \psi_1^2 \right)_{,ij} 
  + \frac{3\eta^4}{200} \nab^{-2} \nab^{-2} D_{ij} \left( \nab^2 \psi_{1,k} \nab^2 \psi_{1}^{~,k}   -  \psi_{1,klm} \psi_{1,}^{~klm} \right)
\nonumber \\
   &+  \frac{\eta^4}{700} \nab^{-2} \left[  10 \psi_{1,kl} \psi_1^{,kl} - 3 (\nab^2 \psi_1)^2 
\right]_{,ij} \,, \label{sol:gamma} 
\end{align}
where in the last step we have made use of the operator
\be
  D_{ij} = \partial_i \partial_j - \frac{\delta_{ij}}{3} \nab^2  \,.
\ee

To get the second-order displacement field, we first note
that the spatial metric, here for scalar perturbations only but to arbitrary order, can be decomposed as
\be \label{decomp}
 \gamma_{ij} := \frac{\delta_{ij}}{3} \hat \gamma + D_{ij} \hat \gamma^\parallel \,,
\ee
where $\hat \gamma = \hat \gamma_1 + \hat \gamma_2/2+\ldots$ is just the trace of $\gamma_{ij}$,
whereas $\hat \gamma^\parallel = \hat \gamma^\parallel_1 + \hat \gamma^\parallel_2/2 +\ldots$ is its
longitudinal part. Applying on this definition the $D^{ij}$ operator then gives the
``longitudinal mode extractor'' 
\be \label{extract}
  \hat \gamma^\parallel  =  \frac 3 2 \nabla^{-2} \nabla^{-2} D^{ij} \gamma_{ij} \,.
\ee
Now, applying  $\frac 3 2 \nabla^{-2} \nabla^{-2} D^{ij}$  on both sides of~(\ref{gamma:displacement}), it is straightforward to obtain an expression for the Lagrangian displacement field (valid only up to second order)
\be \label{Fsecond}
  {\cal F} = \frac 3 4 \nab^{-2} \nab^{-2} D^{ij} \left( \gamma_{ij} + 4 {\cal F}_{,ij} {\cal B}  - {\cal F}_{,il} {\cal F}^{,l}_{,j} \right) \,,
\ee
which can be easily solved with the conventional Ansatz ${\cal F} = {\cal F}_1 + {\cal F}_2/2$, i.e., order by order. 
Restricting for the moment to the fastest growing mode and using the result for $\gamma_{ij}$, Eq.\,(\ref{sol:gamma}),
we then obtain for the scalar part of the displacement field, up to second order
\be \label{displacement}
  {\cal F}^+ = - \frac{\eta^2}{10} \psi_1 - \frac{3 \eta^4}{700} \nab^{-2} \mu_2  - \frac{3 \eta^2}{10} \Theta_0 - \frac{\eta^2}{5} \psi_1^2 \,,
\ee
with the spatial kernels $\Theta_0$ and $\mu_2$ given in Eqs.\,(\ref{Theta0})--(\ref{mu2}). 
This result agrees with Ref.~\cite{Rampf:2014mga} for their $f_{\rm nl}=-5/3$,
  agrees with Ref.~\cite{Villa:2015ppa} for their $a_{\rm nl}=0$. Note that in order to arrive at the expression \eqref{Fsecond} and subsequent results we have again used the identity in Eq.~\eqref{identity}.

The first two terms on the RHS of Eq.~(\ref{displacement}) are the
well-known Newtonian parts of the displacement field up to second
order, whereas the latter terms are relativistic corrections.  
Finally, we note that the resulting
relativistic coordinate transformation
\begin{align}
\label{transf:x}
 \begin{split}
  \eta(\eta,\fett{q})  &= \eta \\
  x^i(\eta,\fett{q}) &= q^i - \frac{\eta^2}{10} \psi_1^{,i} - \frac{3 \eta^4}{700} \nab^{-2} \mu_2^{,i}  - \frac{3 \eta^2}{10} \Theta_0^{,i} - \frac{2\eta^2}{5} \psi_1 \psi_1^{,i} 
 \end{split}
\end{align}
is the 4D gauge transformation from the
synchronous-comoving gauge to the total matter gauge (which also makes use of the proper time coordinate), as already noted in
Refs.\,\cite{Rampf:2014mga,Villa:2015ppa}. Equation~\eqref{transf:x} denotes the relativistic trajectory of fluid elements on constant time-like hypersurfaces up to second order in cosmological perturbation theory,  including, however, only the fastest growing modes. In the following section we will show how to incorporate also the decaying modes.

\subsubsection{Growing and decaying mode}

Substituting into Eq.~(\ref{Fsecond}) the general solutions including decaying
modes, and using the same techniques as described in the previous
section, we obtain for the displacement up to second order (i.e.,
including the first-order results)
\begin{align}
\label{F2:decay}
{\cal F} &= -\frac{\eta_{\rm ini}}{3}\sigma_{1{\rm ini}}\etaonetai^{-3} -\frac{\eta^2}{10}\psi_1 \Big[1+\frac{2}{3}\etaonetai^{-5}\Big]
 - \frac{3 \eta^4}{700} \nab^{-2} \mu_2  - \frac{3 \eta^2}{10} \Theta_0 - \frac{\eta^2}{5} \psi_1^2 \nonumber \\
 &+\etaini^4 \etaonetai^{-1} \nabla^{-2} \left(  - \frac{1}{40} \nabla^{-2} D^{kl} \psi_{1,km} \psi_{1,l}^{,m} + \frac{1}{100} \psi_{1,lm} \psi_{1}^{,lm} + \frac{1}{100} [\nabla^2\psi_{1}]^2 \right)  \nonumber \\
 &+ \etaini^4 \etaonetai^{-6} \nabla^{-2} \Bigg( \frac{1}{240} \nabla^{-2} D^{kl} \psi_{1,km} \psi_{1,l}^{,m} + \frac{1}{800} \psi_{1,lm} \psi_{1}^{,lm}  + \frac{1}{800} [\nabla^2\psi_{1}]^2  \nonumber \\ 
  &\qquad + \frac 9 8 \left[ 5 \nabla^{-2} (\zeta_1+\psi_1)_{,kl} \nabla^{-2} (\zeta_1+\psi_1)^{,kl} - (\zeta_1+\psi_1)^2  \right]  \Bigg)  \nonumber \\
 &+ \frac{27}{4} \etaonetai^{-6} \nabla^{-2}  \left[ (\zeta_1+\psi_1)_{,k} (\zeta_1+\psi_1)^{,k} - \nabla^{-2}(\zeta_1+\psi_1)_{,ijk} \nabla^{-2}(\zeta_1+\psi_1)^{,ijk}   \right] \nonumber \\
 &+ \etaini^2 \etaonetai^{-1} \nabla^{-2} \Bigg(  -\frac{9}{20} \nabla^{-2} \left[ (\zeta_1+\psi_1)_{,k} \nabla^2 \psi_{1}^{,k} + \nabla^{-2} (\zeta_1+\psi_1)_{,ijk} \psi_1^{,ijk}  \right]    \nonumber \\
 &\qquad -6 \psi_{1,kl} \nabla^{-2} (\zeta_1+\psi_1)^{,kl} -\frac{3}{10} (\zeta_1+\psi_1) \nabla^2 \psi_1  \Bigg) \nonumber \\
 &+  \etaini^2 \etaonetai^{-6} \nabla^{-2} \Bigg(  \frac 1 5 \nabla^{-2} \left[  (\zeta_1+ \psi_1 )_{,k} \nabla^2 \psi_{1}^{,k} + \nabla^{-2} (\zeta_1 +\psi_1)_{,ijk} \psi_1^{,ijk}  \right]   \nonumber \\
 &\qquad + \frac 1 2 \psi_{1,kl} \nabla^{-2} (\zeta_1+\psi_1)^{,kl} + \frac{1}{30} (\zeta_1 + \psi_1) \nabla^2 \psi_1 \Bigg)  \nonumber \\
 &- \frac{\etaini^2}{5} \etaonetai^{-3} \left( \frac{\psi_1^2}{3} + 2\Theta_0 \right)
   - \nabla^{-2} \nabla^{-2} D^{ij} \Bigg(  \etaini \etaonetai^{-3} \sigma_{{\rm ini},ij} \psi_1   \nonumber \\
 &\qquad   + \frac{\etaini^2}{12} \etaonetai^{-6}  \sigma_{{\rm ini},il} \sigma_{{\rm ini},j}^{,l} 
+ \frac{\etaini^3}{40} \etaonetai^{-1}  \left\{ \sigma_{{\rm ini},il} \psi_{1,j}^{,l}+\sigma_{{\rm ini},jl} \psi_{1,i}^{,l} \right\}  \left[   1 + \frac 2 3  \etaonetai^{-5} \right]  \Bigg)  \,.
\end{align}
This is our main result. Only a few terms in this expression, namely
the fastest growing modes, have been reported in the literature. All
terms including decaying modes have not been reported before. The relativistic trajectory, including also the decaying modes, can then be easily obtained by observing that $x^i= q^i + {\cal F}^{,i}$, where the scalar ${\cal F}$ is given in Eq.~\eqref{F2:decay}, and $q^i$ denote the initial position of the fluid elements.

\section{Discussion and conclusions}
\label{sec:discussion}

In this paper we have derived the Lagrangian
displacement field using second order relativistic cosmological
perturbation theory, working in synchronous-comoving gauge and using
gauge-invariant scalar perturbations. For simplicity we restricted our
derivation to an Einstein--de Sitter universe, although the
calculations can be readily extended to other models of the Universe.  Here we have focused on determining all the
terms in the solutions, including the decaying modes of the
displacement field, which were so far unknown and could become
important at early times, where the influence of the
cosmological constant should be negligible. For the fastest growing
mode of the displacement field, we recover known results from the literature,
that is, the most dominant contribution to the displacement comes from
the well-known Newtonian part, whereas the fastest growing mode of the
general relativistic corrections to this displacement should become
relevant to cosmological structure formation only on very large
scales.

For the decaying modes, the impact of the general relativistic
corrections is difficult to estimate by analytical means, however it
is expected that these corrections could play an important role at
early times --- on a wealth of scales. This expectation should be
tested by using our novel displacement field for generating initial
conditions for $N$-body simulations, and then analysing the impact of
these \emph{relativistic transients} on the late-time gravitational
dynamics in comparison with standard methods used in the literature
(e.g., 2LPT).  For a possible practical implementation of our
expressions, existing algorithms for generating initial conditions
(e.g., Refs.~\cite{Scoccimarro:1997gr,Crocce:2006ve}) could be modified to
include the relativistic corrections.

Finally, we note that our results allow us write down the
synchronous-comoving metric for scalar perturbations up to second
order, including decaying modes, which also has not been published
before. We did not write down this expression explicitly, but using
$\psi_2$ and $E_2$ given in Section \ref{grow_sol_sect}, it is
straightforward to obtain the corresponding metric for growing and
decaying modes.\\

There are many avenues to extend the results reported in this paper in
future work. To set the initial conditions at later times
it will be useful to use a $\Lambda$CDM background model. This will also require the use of more than one fluid, and allowing for pressure (as done, for example, in \cite{Christopherson:2012kw} at first order), to get a realistic model. In this work we only used scalar perturbations at first and second order. Another extension will be to include contributions from
vector and tensor perturbations (in the case of vectors at first
\emph{and} second order, since in multi-fluid systems second order
vector perturbations will be generated inevitably, \cite{Christopherson:2010dw}
This will result in smaller corrections since the contribution
are supposed to be sub-dominant compared to scalars, at least on large
scales, but might nevertheless have observable effects. Gravity is
after all nonlinear.

\section*{Acknowledgements}

The authors would like to thank Pedro Carrilho for useful discussions.
KAM is supported, in part, by STFC grant ST/J001546/1, JCH acknowledges support from PAPIIT-UNAM grants IA-101414 and IN-103413-3, as well as SEP-CONACYT grant No.
239639. CR
acknowledges the support of the individual fellowship RA 2523/1-1 from
the German research organisation (DFG). This work is supported in part
by the U.S.\ Department of Energy under Grant No. DE-FG02-97ER41029.

\appendix

\section{Solutions in Fourier space}\label{app:Fourier}

In Fourier space the solution to \eq{laplacian:psi2} for $\psi_2$ can
be rewritten in terms of the growing mode solutions
\bea
\label{fourier:psi2}
\widetilde{\psi_{2}(\bk)} = 
\int\int\frac{d^3k_1d^3k_2}{(2\pi)^3} \delta_D(\bk_{\rm q})\psi_1(\bk_1)\psi_1(\bk_2) 
\Bigg\{
 \frac{\eta^2}{10}\Bigg[\frac{k_1^2k_2^2}{k^2} &+& \frac{(k_1^2+k_2^2)\bk_1\cdot\bk_2}{k^2} 
+ \frac{(\bk_1\cdot\bk_2)^2}{k^2} \Bigg] \nonumber \\
&+&\frac{\eta^4}{200}\left[\frac{(k_1^2k_2^2)\bk_1\cdot\bk_2}{k^2} - \frac{(\bk_1\cdot\bk_2)^3}{k^2}\right]
\Bigg\}\,,
\eea
where $\bk,\,\bk_1$ and $\bk_2$ are independent vectors in Fourier space. $\bk_{\rm q}\equiv \bk - \bk_1 - \bk_2$, and the Dirac delta $\delta_D(\bk_{\rm q})$ guarantees momentum conservation. Also we have symmetrised over $\bk_1$ and $\bk_2$. Similarly, the real-space solution for $E_2$ (Eq.\,(\ref{solE2})) is easily transformed to Fourier space:
\be
E_2=
\int\int\frac{d^3k_1d^3k_2}{(2\pi)^3} \delta_D(\bk_{\rm q})\frac{1}{k^2}\left[
\frac{1}{10}\mathfrak{K}_1\eta^2+\frac{1}{28}\mathfrak{K}_2\eta^4
\right]\psi_1(\bk_1)\psi_1(\bk_2)\,.
\ee

where
\begin{align}
\mathfrak{K}_1=& -\Bigg[2\left(k_1^2+k_2^2 \right)+3\bk_1\cdot\bk_2\Bigg]   
-\frac{3}{k^2}\Bigg[k_1^2k_2^2+\left(k_1^2+k_2^2 \right)\bk_1\cdot\bk_2
+\left(\bk_1\cdot\bk_2\right)^2
\Bigg]
\,, 
\end{align}
\begin{align}
\mathfrak{K}_2=&-\Bigg[\frac{20}{50}\left(\bk_1\cdot\bk_2\right)^2-\frac{6}{50}k_1^2k_2^2\Bigg] 
+\frac{3}{k^2}\bk_1\cdot\bk_2\Bigg[-\frac{7}{50}\ k_1^2k_2^2+\frac{7}{50}\left(\bk_1\cdot\bk_2\right)^2
\Bigg]
\,,  
\end{align}
and we again symmetrised over $\bk_1$ and $\bk_2$.

For the second-order solution of the density contrast in Fourier space 
we get, using Eq.\,(\ref{delta2+}),
\be
{\delta_2} = 
\int\int\frac{d^3k_1d^3k_2}{(2\pi)^3} \,\delta_D(\bk_{\rm q}) \Big\{
 -4 (k_1^2 + k_2^2) - 3\bk_1\cdot\bk_2 
+ \frac{\eta^2}{10}\left[\frac{10}{7} k_1^2k_2^2 
+\frac{4}{7} (\bk_1\cdot\bk_2 )^2\right] \Big\}
\frac{\eta^2}{10} \psi_1(\bk_1)\psi_1(\bk_2)\,.
\ee
This result is in agreement with Eqs.\,(94) and~(A10) of Ref.\,\cite{Bruni:2013qta}.
\\

\end{document}